\documentclass{midl} 

\usepackage{url}
\usepackage{colortbl}
\usepackage{ulem}
\usepackage{mwe} 
\jmlrvolume{}
\jmlryear{2022}
\jmlrworkshop{}
\editors{}

\title[Give me a knee radiograph]{Give me a knee radiograph, I will tell you where the knee joint area is: a deep convolutional neural network adventure}

\midlauthor{\Name{Shi Yan\nametag{$^{1}$}} \Email{}\\
\Name{Taghi Ramazanian\midlotherjointauthor\nametag{$^{2}$}} \Email{}\\
\Name{Elham Sagheb\midlotherjointauthor\nametag{$^{1}$}} \Email{}\\
\Name{Walter K. Kremers\midlotherjointauthor\nametag{$^{2}$}} \Email{}\\
\Name{Vipin Chaudhary\nametag{$^{3}$}} \Email{}\\
\Name{Michael Taunton\nametag{$^{4}$}} \Email{}\\
\Name{Hilal Maradit Kremers\midljointauthortext{corresponding authors}\nametag{$^{2}$}} \Email{}\\
\Name{Ahmad P. Tafti$^{*}$\nametag{$^{5, 6}$}} \Email{}\\ \AND
\addr $^{1}$Department of Artificial Intelligence and Informatics, Mayo Clinic, Rochester, MN, United States
\addr $^{2}$Department of Health Sciences Research, Mayo Clinic, Rochester, MN,  United States \\
\addr $^{3}$Department of Computer and Data Sciences, Case Western Reserve University, Cleveland, OH, United States \\
\addr $^{4}$ Department of Orthopedics, Mayo Clinic, Rochester, MN,  United States \\
\addr $^{5}$ Department of Computer Science, University of Southern Maine, Portland, ME, United States \\
\addr $^{6}$ Dubyak Center for Digital Science and Innovation, University of Southern Maine, Portland, ME, United States
}

\begin{document}

\maketitle

\begin{abstract}

Knee pain is undoubtedly the most common musculoskeletal symptom that impairs quality of life, confines mobility and functionality across all ages. Knee pain is clinically evaluated by routine radiographs, where the widespread adoption of radiographic images and their availability at low cost, make them the principle component in the assessment of knee pain and knee pathologies, such as arthritis, trauma, and sport injuries. However, interpretation of the knee radiographs is still highly subjective, and overlapping structures within the radiographs and the large volume of images needing to be analyzed on a daily basis, make interpretation challenging for both naive and experienced practitioners. There is thus a need to implement an artificial intelligence strategy to objectively and automatically interpret knee radiographs, facilitating triage of abnormal radiographs in a timely fashion. The current work proposes an accurate and effective pipeline for autonomous detection, localization, and classification of knee joint area in plain radiographs combining the You Only Look Once (YOLO v3) deep convolutional neural network with a large and fully-annotated knee radiographs dataset. The present work is expected to stimulate more interest from the deep learning computer vision community to this pragmatic and clinical application.


\end{abstract}

\begin{keywords}
Deep Learning, Medical Imaging, Knee Radiographs, Knee Joint Area Localization, Knee Joint Area Classification.  
\end{keywords}

\section{Introduction}

Knee pain is one of the most common symptoms affecting approximately 25\% of adults, limits function, mobility, and impairs quality of life  \citep{grotle2008prevalence, jinks2002measuring, peat2001knee, felson1998update}. While knee osteoarthritis (OA) is the most common cause of knee pain in people 50 years or older  \citep{zeni2010clinical}, knee injuries are another common reason for knee pain, especially in young and physically active adults \citep{gage2012epidemiology}. Treatment options for knee pain in patients with knee OA include medical and surgical interventions, such as total knee arthroplasty (TKA). TKA is one of the most common inpatient procedures  \citep{kurtz2014impact}. Over 500,000 TKA procedures are performed in the United States annually, and approximately 4.7 million Americans are currently living with TKA implants \citep{kremers2015prevalence}. Knee radiographs are not only standard care of TKA follow-up \citep{sloan2018projected, kunutsor2017systematic, kremers2013determinants}, but also are the most common imaging modality used in the emergency departments (EDs) for evaluation of the knee joint. Although most physicians can clinically distinguish injuries like fractures, failure to obtain an adequate history, and expectations on the part of patients lead them to often order radiographs for most patients with acute knee  injury \citep{long1985radiographic, brand1982protocol, hall1976overutilization}. Consequently, they are the workhorse  of knee imaging, and almost every symptom or sign is initially evaluated by an x-ray. Furthermore, radiographs provide very invaluable information across the entire spectrum of knee pathology, including arthritis, trauma, oncology, sports injuries, metabolic disease, and TKA complications \citep{math2006imaging}. Although knee injuries  are treated by a wide range of clinicians, an estimated 6.6 million knee injuries presented to U.S. emergency departments from 1999 through 2008 \citep{gage2012epidemiology}. In summary, knee injuries are not only very common in youth and young adults,  radiographs for knee pain are increasingly common in the elderly \citep{gage2012epidemiology}.

In recent years, several efforts are devoted to building computational mechanisms to understand and interpret knee radiographs in an automatic fashion \citep{paul2019automated, gorriz2019assessing, liu2019deep, paul2019automated, tiulpin2018automatic, antony2017automatic, antony2016quantifying, thomson2015automated, woloszynski2010signature, shamir2009early}. Automatic localization of the region of interest (ROI) which clearly points to knee joint area has played a vital role in almost all of these studies, where it becomes the very first essential step in knee x-ray analysis. The impact of automatic detection and precise localization of knee joint area will be significant for patient care when considering large daily volume of x-rays, and only about 15\% result in clinically significant findings \citep{tandeter1999acute, saxena1992role}. The rest thus adds substantial costs to both the patients and the health care system. Furthermore, subjective analysis of the large-scale knee radiographs often comes with considerable variability by overburdened radiologists and practitioners. 

While  artificial intelligence (AI) strategies and computational mechanisms have  been  around  for  several years  in  the  literature, the  use  of  advanced machine learning methods in  knee joint area localization and classification on plain radiographs  has been  limited  to  few  studies \citep{tiulpin2017novel, antony2016quantifying, shamir2009early}. These studies can be categorized into three different methods, including template matching, sliding-window, and machine learning-based scoring algorithms. The motivation of our study is therefore to revisit the state of the art in the literature, and provide better and more accurate solutions to fulfill the following objectives: (1) creating a practical and consistent annotation guideline to create a fully-annotated knee radiography dataset that covers diverse types of knee radiographs, including knee anteroposterior (AP) view, knee lateral view, total knee arthorplasty (TKA) AP view, and TKA lateral view; (2) developing explainable deep convolutional neural network for precise detection, localization, and classification of knee joint areas; and (3) drawing attention from artificial intelligence community and deep learning computational vision to this pragmatic application, opening doors for several attractive directions to pursue deep learning in orthopedics research.

We briefly summarize the {\bf main significance} of our work as follows:

\subsection{Clinical Significance}
With growing number of TKA patients, the routine radiographic follow-up remains an over- whelming task for most orthopedic centers. There is also an indispensable need for evaluation of the knee radiographs in patients with different knee bony pathology which is beyond the expertise of most health care providers. The need for evaluation of a large volume of knee x-ray images on a daily basis creates a burden for both naive and experienced practitioners, and may lead to wrong interpretation or missed diagnoses. Our system, which is basically considered as an underlying AI-powered component for a wide range of knee radiographic analysis, aims to automatically and objectively localize and classify the knee joint area. Here, we developed a new technological solution that can detect, localize, and classify the knee joint area with high degree of accuracy, which could be a substantial advantage in different clinical settings, including TKA, knee injuries, arthritis, and more.


\subsection{Technical Significance}

First, we developed an annotation guideline that fits well in annotating knee radiographs, particularly for localization and classification of knee joint area in plain radiographs. Second, we computationally established a large-scale fully-annotated knee radiographic dataset to train and validate a data hungry deep learning method for autonomous localization and classification of knee joint area. Third, the present work is the first to utilize deep convolutional neural networks in the problem statement, trying to automatically learn the features for diverse types of knee radiographs, including AP and lateral view plus those with and without TKA, replacing the traditional methods of analyzing handcrafted features with new state-of-the-art deep learning strategies. Our implementation used the YOLO (You Only Look Once)-based deep neural networks to detect, localize, and classify knee joint area in x-rays, simultaneously predicting multiple bounding boxes along with class probabilities. Fourth, we built a large-scale knee radiographic dataset by integrating publicly available data with privately Mayo Clinic x-rays, which has advantages of covering more diverse and large volume of the images compared to the previous studies with small or moderate size publicly available datasets (e.g., OAI, MOST) \citep{tiulpin2017novel, antony2016quantifying, shamir2009early}. Finally, this work is the first that can tackle the knee joint area localization within different types of knee radiographs, including AP and lateral view, with or without TKA. 

The rest of the paper is organized as follows. We will explain the materials and methods in the next section. Experimental validation of the method and scientific visualization are demonstrated in Section 3. Section 4 further discusses the work, and it draws conclusion and outlook. 

\section{Materials and Methods}

In this  section, we  shall  begin  with  the  dataset  description, and then explain  the  pipeline  and the underlying computational components for detection, localization, and classification of the knee joint area in x-ray images. 
\begin{figure}[htbp]
\floatconts
  {fig:7Types}
  {\caption{A sample subset of the knee x-ray images available in the proposed integrated dataset. This figure shows anteroposterior (AP) and Lateral views of one or both (Bilateral) knee joints with and without total knee arthroplasty (TKA). (a) AP view of the knee joint, (b) Lateral view of the knee joint, (c) AP view of Bilateral (right and left) knee joints, (d) AP view of the knee joint with total knee arthroplasty (TKA), (e) Lateral view of the knee joint with total knee arthroplasty (TKA), (f) AP view of the both (Bilateral) knee joints with and without TKA, (g) AP view of the both (Bilateral) knee joints with TKA. }}
  {\includegraphics[width=0.8\linewidth]{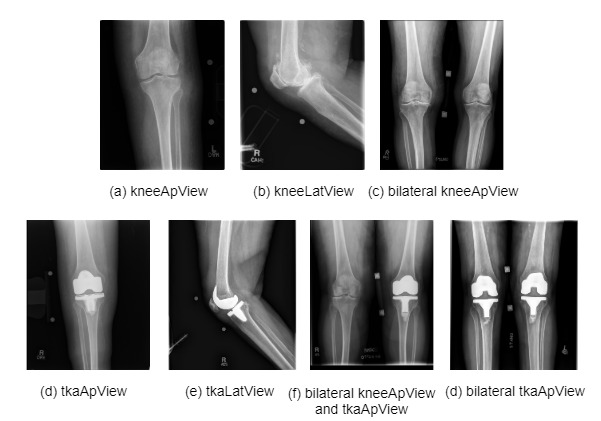}}
\end{figure}

\subsection{Dataset Collection and Annotation}
Following IRB approval, we conduct this study by integrating a publicly available knee radiographic dataset, namely the Osteoarthritis Initiative \citep{OAIDataset}, with Mayo Clinic knee radiographs. This helps to create a diverse training data, better managing biases in data selection when we are building the deep learning-powered localization and classification model. With the help of an orthopedic surgeon, our team developed an annotation guideline for manually localization and classification of knee joint area within the following types of knee radiographs: 1) knee AP view; 2) knee lateral view; 3) TKA AP view; and 4) TKA lateral view. The guideline instructed the annotators to label the knee joint areas as ``kneeApView", ``kneeLatView", ``tkaApView", and ``tkaLatView", indicating knee AP view, knee lateral view, TKA AP view, and TKA lateral view, respectively. The annotation was performed by a trained annotator, and it resulted to obtain an excellent Intersection over Union (IoU) score \citep{rezatofighi2019generalized, rahman2016optimizing} of 0.954 for localization in average, within the first and second batch, including 70 and 100 knee radiographs. The average F1-Score for knee joint area classification was 1.000 among those two batches.  

From 900 randomly selected OAI knee radiographs, 655 images were annotated. We also added 5,465 out of 6,372 randomly selected Mayo Clinic knee radiographs based on the inclusion criteria, in which it resulted to total of 6,120 knee radiographs for annotation. \figureref{fig:7Types} presents a sample subset of all types of the knee radiographs exist in our integrated dataset. Out of 6,120 knee radiographs, we
ended up with 8,634 fully-annotated knee joint areas to train and validate the deep learning localization and classification model. A sample subset of these four classes, including  kneeApView, tkaApView, kneeLatView, and tkaLatView are presented in \figureref{fig:4Classes}. \tableref{tab:categoryInfo} shows further details on the number and types of the knee joint areas within our dataset. While the OAI dataset did not represent gender-specific information, the Mayo Clinic radiographs provided gender-specific disparity of 184 females and 219 males that made us confident regarding gender diversity in our integrated dataset.


\begin{table}[htbp]
\floatconts
  {tab:categoryInfo}%
  {\caption{8,634 fully-annotated knee joint areas have been provided by the annotators to train, validate, and test the deep learning localization and classification model.}}%
  {\begin{tabular}{ll}
  \bfseries Knee joint area category &  \bfseries No. (\%)\\
  kneeApView & 3,608 (41.78)\\
  kneeLatView  & 1,256(14.55)\\
  tkaApView  & 2,121 (24.57) \\
  tkaLatView  & 1,649 (19.10) \\
  {\bf Total}  & {\bf 8,634}
  \end{tabular}}
  \end{table}

\subsection{Deep CNN for Detection, Localization, and Classification of Knee Joint Area in Plain Radiographs}

The method used in this paper is mainly based on YOLO v3 \cite{redmon2018yolov3, PyTorchYOLOv3} deep convolutional neural network. The YOLO v3 is equipped with four losses as: (1) Mean Squared Error (MSE) of $x, y, w, h$ between label and the detection result; (2) Binary Cross Entropy (BCE) of objectness score; (3) BCE of no objectness score; and (4) BCE of multi-class prediction results. Inspired by the implementation at \citep{PyTorchYOLOv3}, we replaced MSE of $x, y, w, h$ with Generalized Intersection over Union (GIoU) \cite{Rezatofighi_2018_CVPR, rezatofighi2019generalized}, which is basically considered as the most widely-used evaluation metric for object detection. However, when there is no intersection between the ground-truth object detection and the prediction one, the IoU then will be equal to 0, in a way that means there is no a gradient when we utilize it within the loss function. GIoU can tackle the current problem by it differentiability attribute. GIoU is calculated as follows, where A and B are the prediction and the ground truth respectively, C is the smallest convex hull that include both A and B.

$$
G I o U=\frac{|A \cap B|}{|A \cup B|}-\frac{|C|(A \cup B) |}{|C|}=\operatorname{IoU}-\frac{|C \backslash(A \cup B)|}{|C|}
$$



\begin{figure}[htbp]
\floatconts
  {fig:4Classes}
  {\caption{Knee joint area classes. This multi-class classification includes four classes as: kneeApView, tkaApView, kneeLatView, and tkaLatView.}}
  {\includegraphics[width=1\linewidth]{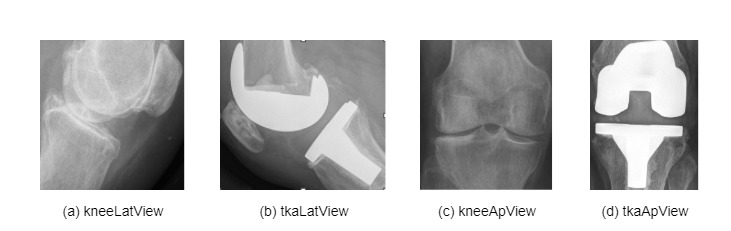}}
\end{figure}

We will first briefly introduce You Only Look Once (YOLO) \cite{redmon2016you} deep neural network. The main idea of YOLO is by given an image, the network will predict the location of the bounding box(es) around object(s), the class probability of the object(s), and the confidence, which means whether the bounding box(es) contains the object(s) or not. The location of a bounding box is represented by $(x,y,w,h)$, where $(x, y)$ is the center of the bounding box, and $(w,h)$ is the width and height of the bounding box. And $(x,y,w,h)$ is a relative value divided by the real width and height of the original image. For every single image, YOLO will divide it into $S \times S$ grids, and for each grid, YOLO will predict $B$ bounding boxes, the bounding boxes contain the location and confidence. We can treat them as a $(B \times 5+C)$ tensor, where $C$ represents the number of classes. Finally, for each image, YOLO will generate a tensor with the dimension of $S \times S \times (B\times5+C)$. In summary, the YOLO system can be visualized as in \figureref{fig:yoloStep}.

\begin{figure}[htbp]
\floatconts
  {fig:yoloStep}
  {\caption{Understanding of the YOLO system. The basic idea of YOLO is: (1) divide image into $S \times S$ grids, (2) predict bounding boxes for each grid, (3) after the filter step, get class probability, (4) draw bounding boxes and provide the labels.}}
  {\includegraphics[width=.96\linewidth]{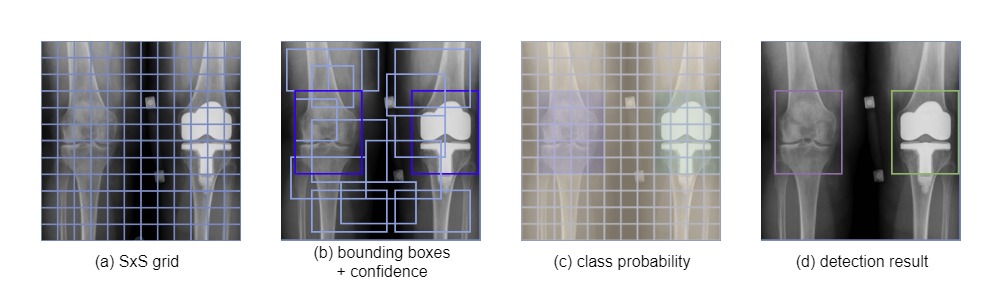}}
\end{figure}
YOLO is a widely-used computational method for real-time object detection, but it also comes with some drawbacks. Since softmax activation function is used in YOLO, thus it cannot deal with the problem of multi-label classification. In YOLO v3, logistic activation function replaced the softmax function to solve the problem. This paper does not have the space to go deeper into the details of YOLO deep convolutional neural network; interested readers are therefore referred to \cite{redmon2016you, redmon2017yolo9000, redmon2018yolov3} for further reading.

\section{Experimental Validation and Scientific Visualization}

Several experiments were performed to measure and examine the performance of the model. We shall begin with introducing the test bed, train, and validation subsets. 

From the computational perspective, a high performance cluster equipped with 2 Tesla V-100 GPUs with 6.21G memory in total was used to run the model. All implementations were made using PyTorch. After several experiments to find the best tuned hyper-parameters, we did set the epochs to 48, batch size 32, and use Adam with learning rate of 0.001. The IoU threshold was configured as 0.5. Regarding the training and validation dataset, we did first split the Mayo Clinic x-rays based on gender, then for both genders, we divided the data based on patient ID. After that, we used 5-fold cross validation to generate train, validation and test data. 

To evaluate the performance of the deep learning model, we accounted for 5-fold cross validation fashion to train and validate the knee joint area localization and classification model. Then we selected the best model to train on the whole dataset but test dataset, then detect the knee joint area on our test dataset. Finally we got a IOU score with 0.941 on test dataset, which illustrates that our method can lead the machine to achieve an excellent agreement with the annotators. \figureref{fig:detectionResults}  visually demonstrates some samples for detection, localization, and classification results. One can see that the deep learning model and our annotation guideline both contribute to the excellent IoU score, which also indicates the reliability and stability of our annotation guideline. 

\begin{figure}[htbp]
\floatconts
  {fig:detectionResults}
  {\caption{A sample subset of the detection, localization, and classification, to visually demonstrate the results. The score on the image is the confidence score.}}
  {\includegraphics[width=0.96\linewidth]{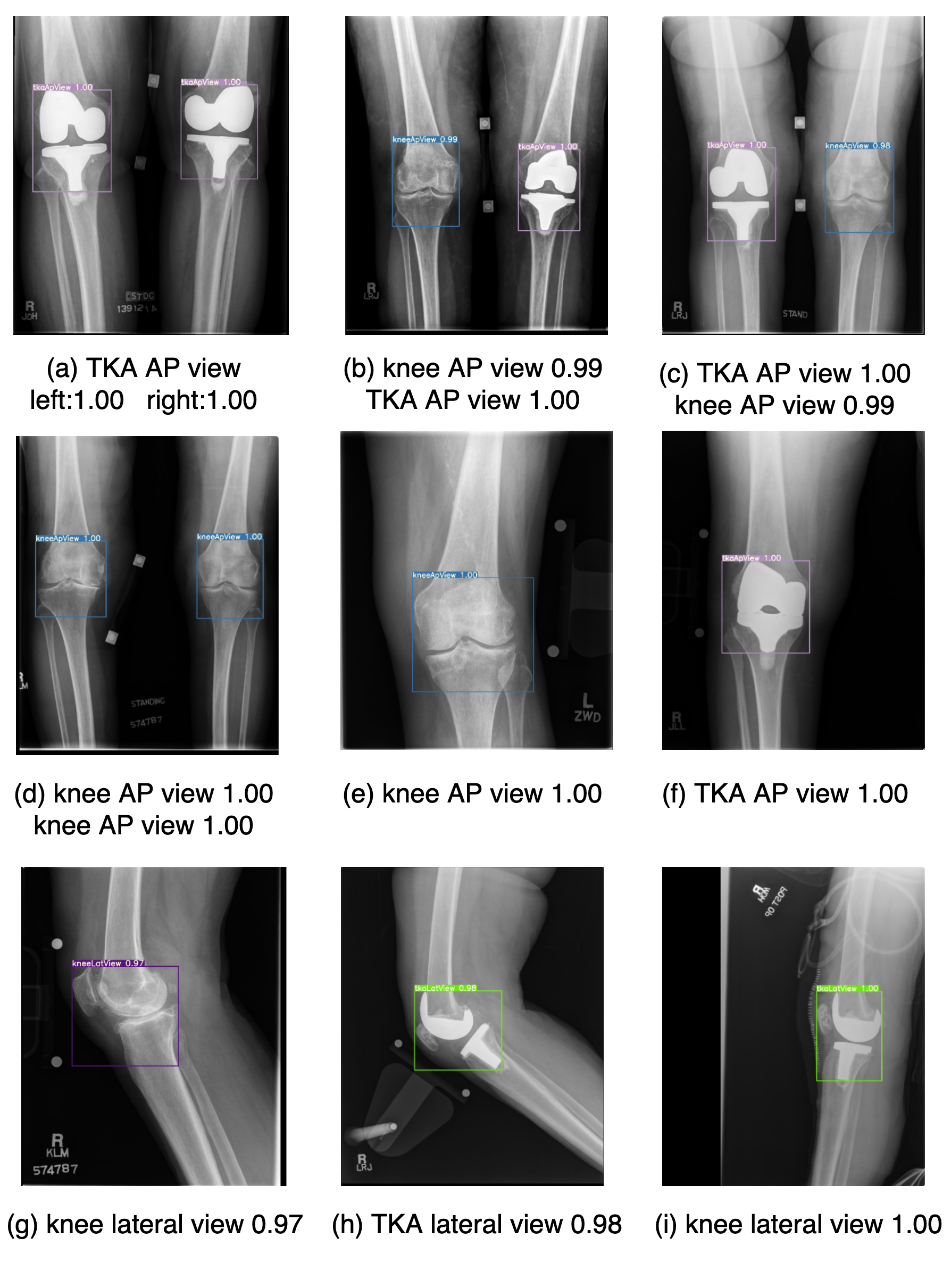}}

\end{figure}


Regarding the classification task, as illustrated in \tableref{tab:results}, we finally obtained the precision, recall, map@0.5, and F-1 score of 0.990, 0.998, 0.993, and 0.994, respectively on test dataset, which indicate excellent performance.

 

\begin{table}[htbp]
\floatconts
  {tab:results}%
  {\caption{The performance across all types of the knee joint area on test dataset.}}%
  {\begin{tabular}{lllll}
  \bfseries Class & \bfseries Precision & \bfseries Recall & \bfseries mAP@0.5 & \bfseries F-1 Score\\
  all         & 0.990 & 0.998 & 0.993 & 0.994\\
  kneeApView  & 1.000 & 1.000 & 0.995 & 1.000\\
  tkaApView   & 0.991 & 1.000 & 0.995 & 0.995\\
  kneeLatView & 1.000 & 0.993 & 0.995 & 0.996\\
  tkaLatView  & 0.971 & 1.000 & 0.988 & 0.985
  \end{tabular}}
\end{table}

\section{Discussion, Conclusion, and Outlook}

In this work, we presented a YOLO v3-based deep neural network architecture to precisely detect, localize, and classify knee joint area in different types of knee radiographs. Our work carries a list of novelties as follows: 

\begin{itemize}
  
    \item {\bf Deep learning computational vision:} Previous publications in the literature only employed conventional machine learning components with handcrafting imaging features. The handcrafted designed features are often over-specified or incomplete, and they take a long time to design and validate. In contrast to the previous works, this study utilizes a deep convolutional neural network strategy with a large-scale training data to automatically and objectively learn the diverse features of different types of knee radiographs. The proposed deep learning computational vision model has demonstrated excellent performance in the localization and classification of knee joint areas.
    
  \item {\bf Annotation guideline:} We also developed an annotation guideline to manually localize and classify the knee joint area on different types of knee radiographs, with and without TKA. Our annotation guideline resulted in an excellent level of agreement for knee area localization as well as classification between three observers and the deep learning component. The excellent IoU score for localization of the knee area in AP and lateral knee radiographs with and without TKA is one of the unique features of our work.

\item {\bf A large-scale and diverse fully-annotated dataset:} We established a large-scale, diverse, and fully-annotated knee radiography dataset to train and validate the data hungry deep learning method(s).  Noticeable gender-diversity as well as high volume knee radiographs provides distinct advantages over the publicly available datasets \citep{tiulpin2018automatic, antony2017automatic, antony2016quantifying, shamir2009early}.

\item {\bf Clinical Application:} The current computational method has potential for several clinical applications, including detection of different types of pathology around the knee joint and automated radiographic follow-up of TKA patients for complications.  In particular, availability of standardized, reusable, scalable, and transferable deep learning algorithms for automated grading of TJA radiographs can be a paradigm shift in TKA clinical practice. An automated system that can triage abnormal radiographs and automatically measure location of abnormality is a huge productivity advantage. Surgeons can become more efficient by prioritizing patients based on automated radiographic findings, rather than reviewing all radiographs manually. This will improve productivity, freeing up time to focus on value-added activities. Furthermore, information garnered will improve clinical practice by revealing previously unknown surgical technique variability that may significantly influence TKA outcomes.

\end{itemize}

However, some potential limitations should be noted. First, our annotation guideline is built on the anatomical landmarks around the knee joint. Although this provided us an excellent level of agreement in localizing and classifying the knee area, the area itself is confined to a limited window that might interfere with clinical application of the localization. One solution would be resizing the bounding boxes with pre-defined margin to the anatomical landmarks in an automatics fashion. Second, we tested our annotation guideline on AP and lateral view of the knee radiographs. A standard lateral knee radiograph is usually performed in 20º to 30º of knee flexion. Knee radiographs with more flexion obscure the anatomical landmarks we used for annotation; we therefore decided to exclude these types of the knee lateral radiographs due to incompatibility with the annotation guideline.

In conclusion, this study demonstrates a successful application of YOLO v3 deep neural network to reliably and precisely tackle the problem of knee joint area localization and classification in plain radiographs. As part of our future work, we are combining the knee joint areas that are computationally identified from the plain knee radiographs to build further deep learning predictive models with knee pathology, advancing knee imaging informatics research.


\bibliography{midl-fullpaper.bib}

\end{document}